\documentclass [reprint aps,prb,twocolumn,groupedaddress,nofootinbib,floatfix,amsmath,mssymb,prper,floatfix,twocolumn, superscriptaddress]{revtex4-2}
\usepackage{array, makecell}
\usepackage{graphicx} 
\graphicspath{{./Graphics/}}
\usepackage{array,multirow,graphicx}
\usepackage{graphics}
\usepackage{epsfig}
\usepackage{dcolumn}
\usepackage{comment}
\usepackage{amsfonts,color}
\usepackage{appendix}
\usepackage{multibib}
\usepackage[utf8]{inputenc}
\usepackage[table]{xcolor}
\usepackage{bm}
\usepackage{scalerel,amssymb}

\begin{document}
\preprint{APS/123-QED}
\title{Exploring Subfield Interest Development in Undergraduate Physics Students through Social Cognitive Career Theory}
\author{Dina Zohrabi Alaee}
\affiliation{Department of Physics and Engineering Science, Coastal Carolina University, Conway, SC, 29526}
\author{Benjamin M. Zwickl}
\affiliation{School of Physics and Astronomy, Rochester Institute of Technology, Rochester, NY, 14623}

\date{\today}
\begin{abstract}
This study aims to understand how undergraduate physics majors develop an interest in specific subfields. We examine interest formation through the lens of Social Cognitive Career Theory (SCCT) by exploring four key SCCT constructs: learning experiences, self-efficacy, outcome expectations, and proximal environmental influences. We conducted 27 interviews with physics majors across various years of study between 2020 and 2022. Our first research question analyzes SCCT constructs to provide detailed insights into the interest formation of various subfields of physics. Examining these constructs, we better understand the factors influencing students' preferences toward specific subfields. Our findings indicate that positive class experiences and experimental opportunities significantly impacted students' interest in different subfields of physics. This understanding helped us explore significant variations in interest formation between astrophysics and biomedical physics in Research Question 2. By understanding how students decide about their interests, we can provide valuable insights for physics departments and career guidance professionals.

\end{abstract} 
\maketitle
\section{Introduction}
After earning a physics degree, various career paths and specialized fields become available to explore. Approximately 47\% of physics bachelor's graduates find employment in different sectors, 31\% pursue graduate studies in physics or astronomy, and 17\% pursue graduate studies in other fields (e.g., engineering)~\cite{american_institute_of_physics_aip_2020}. 
According to the American Institute of Physics (AIP), condensed matter physics is the most common subfield chosen by physics PhDs, with around 20\% of doctoral candidates specializing in it in the classes of 2019 and 2020~\cite{american_institute_of_physics_aip_2021}. The same AIP report also stated that around 13.5\% of physics PhD candidates specialize in astronomy, astrophysics, or cosmology, while 7.8\% specialize in biophysics. However, it is not always clear what factors influence students' decisions when it comes to choosing a subfield for their graduate studies. Additionally, subfield interest may influence employment an employment search because some subfields have strong links to industry (e.g., optics or materials physics) while other subfields don't (e.g., particle physics). This further underscores the need for better characterization of career decision-making processes, and as the Effective Practices for Physics Programs (EP3)~\cite{mckagan_career_2021} and the TEAM-UP report \cite{national_task} both emphasize, there is a need for improved support and guidance for physics students, as well as the development of effective interventions to support career preparation and decision-making.


Compared to physics programs, engineering and computer science programs typically have stronger connections to specific job opportunities and particular industries, which is often facilitated through internship programs, industry sponsored senior design projects, advisory boards, and accreditation boards. On the other hand, undergraduate physics programs have fewer industry connections, are usually smaller, and have limited resources for non-academic career preparation, such as internship programs. As a result, there is a gap between what physics departments offer and what students may need for successful career paths in physics. 

Our study aims to identify key factors that influence subfield decisions in physics. By understanding different experiences and factors that contribute to positive and negative subfield interests and career decisions, we provide useful insights for physics departments that want to redesign their physics program for career preparation. The research on subfield decision-making is also relevant for efforts to create inclusive educational opportunities and understanding gender differences in fields like physics. These efforts are particularly important in disciplines that aim to attract students from a different range of demographic and disciplinary backgrounds~\cite{house2022quantum}. 

Our study is guided by the following questions:
\begin{itemize}
    \item RQ1 - What factors contribute to students developing interests in specific subfields of physics?
    \item RQ2 - Do interest formation processes differ between different physics subfields, such as biomedical physics and astrophysics?

\end{itemize}
\section{Background}
\begin{figure}[h]
\centering
\includegraphics [trim=3 2 4 5,clip,width=85mm]{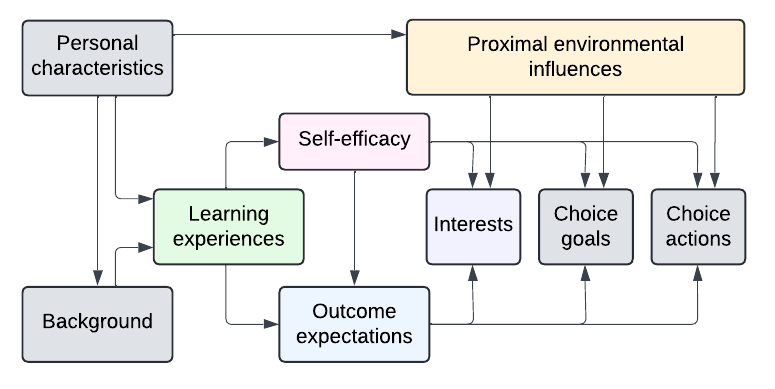}
\centering
\caption{Map of interrelated constructs within Social Cognitive Career Theory. The focus of this study is the non-grey constructs depicted in the figure.}
\label{SCCT}
\end{figure}

We used Social Cognitive Career Theory (SCCT), which describes how specific psychological characteristics or factors impact individuals' career interests in various fields~\cite{lent_toward_1994, lent_longitudinal_2008}. For instance, a qualitative study with undergraduate computer science students employed SCCT to understand how different SCCT factors influence their career interests~\cite{alshahrani2018using}. These constructs are depicted as blocks in Fig.~\ref{SCCT}, illustrating the process of individuals making career-related decisions and choices. For the purpose of this paper, our primary focus is on the colorful (non-gray) constructs depicted in Fig.~\ref{SCCT}, which include: learning experiences, self-efficacy, outcome expectations, interest, and proximal environmental influences. We chose to focus on these constructs from the SCCT in our analysis because these factors were the most frequently mentioned during our interviews with undergraduate students. Additionally, these constructs appeared to the most significantly influenced by the college experiences.

According to the SCCT framework, students' decisions are affected by their pre-college experiences and background influences as well as personal factors such as gender, race/ethnicity, disability status, and socioeconomic status. The learning experiences construct includes all educational activities and opportunities that may affect a student's knowledge, skills, understanding, attitudes, and beliefs related to a particular career path. In SCCT, learning experiences have a direct impact on students' self-efficacy and outcome expectations. Self-efficacy describes a student's belief in their ability to succeed in specific tasks or situations related to their career. Outcome expectations are what people anticipate will happen as a result of performing specific actions, such as pursuing a particular career path. Positive outcome expectations, such as financial rewards, job satisfaction, or career advancement opportunities, can serve as powerful motivators that encourage students to pursue and persist in their chosen careers. Conversely, negative outcome expectations, such as limited job prospects or a poor work-life balance, may demotivate students from pursuing a particular career path. In the SCCT framework, interests give rise to goals and actions. Proximal environmental influences, such as institutional limitations and other factors that extend beyond the classroom setting, can also impact students' interests and goals.

\section{Our prior work on career interest formation}
This study is part of a larger research project aimed at understanding how undergraduate physics students develop interests and make career decisions. Our long-term goal is to help physics departments in preparing students for successful careers. 
Starting in 2020, we have conducted dozens of individual interviews with students majoring in physics to explore their preferred methods in physics (e.g., experimental, computational, or theoretical)~\cite{cardona2021access, alaee2023analyzing} and interests in various subfields of physics~\cite{alaee2022impact, Bennett_2022}.

Interest in specific methods strongly affects students' career paths. The most common private sector jobs are in computing and engineering fields. Over 85\% of physics bachelor's graduates who work in computing fields and more than 50\% of engineers use programming daily~\cite{american_institute_of_physics_aip_2022}. Interest in experimental work often leads to engineering careers or experimental research in graduate school. Across many of the largest subfields of physics, most PhDs are experimental~\cite{american_institute_of_physics_aip_2021-1,american_institute_of_physics_aip_2022}. Low interest in experimental methods can limit career interests in major physics subfields~\cite{alaee2023analyzing}. In our prior work, we found that students less exposed to experimental work in college have lower self-efficacy and interest in experimental physics careers~\cite{cardona2021access}. However, in terms of outcome expectations, many students saw computation as essential for future careers, which increased interest despite lower self-efficacy. Conversely, negative interest in theory often stems from perceiving theoretical work as irrelevant and lacking practical applications~\cite{alaee2023analyzing}. These studies used qualitative inductive approaches to identify categories and subcategories from students' stories about their preferred methods in physics and career interests. 

In our prior work on interest formation in specific subfields, we found that students' interest is shaped by exposure, perceptions, and outcome expectations~\cite{zohrabi2021case, Bennett_2022, alaee2022impact}. Many juniors and seniors are unfamiliar with various physics subfields, revealing a gap between the standard physics curriculum and contemporary areas of physics research. For well-known fields like astronomy and astrophysics, we found that students often had unrealistic expectations about the work involved, leading some to lose interest when they realized it involves desk work and programming rather than on-site telescope observations. Negative experiences in high school biology and chemistry also detered students from subfields overlapping with these sciences, such as biophysics.

\section{Method}
\subsection{Interview Protocol}
We conducted 27 semi-structured individual interviews with college students from different academic years between 2020-2022. The interview protocols were based on the SCCT framework and covered all of the colored SCCT constructs shown in Fig.~\ref{SCCT}. All of the interviews were designed to gain an understanding of their decision-making processes related to physics subfields, while a subset also asked about the methods employed in physics (e.g., experimental, theory, and computation), and intent to pursue graduate school or employment opportunities after graduation. In this paper, we will only focus on findings associated with the subfield of physics that students identified as their interest.


\subsection{Recruitment and Participant Demographics}
We initially interviewed seven physics majors in their third or fourth year of college, who were recruited from summer research programs and an online social platform (Discord) dedicated to physics majors associated with one institution~\cite{cardona2021access}. In the summer and fall of 2020, we expanded our study by interviews with ten physics majors doing summer research at six different institutions across the United States~\cite{alaee2022impact}. 
Finally, during the late summer and fall of 2021, we used various recruitment tools such as emails, posters, and Discord messages to recruit ten more participants from a single institution, many of whom were not participating in undergraduate research. Table~\ref{Tab: demographic} summarizes the participants' demographic characteristics, and Table~\ref{tab: subfield} briefly outlines their subfield of interest. For brevity, we use the label astrophysics to encompass both astronomy and astrophysics and the label biomedical physics to encompass biophysics, medical physics, and nuclear medicine.


\begin{table*}[tbh]

    \caption{Demographic characteristics of participants\label{Tab: demographic}}
    \label{tab:1}
    
\begin{tabular}{c|c|c|c|c|c|c|c|c|c|c|c|c}
\hline
\multicolumn{1}{c|}{Data collection} & \multicolumn{3}{c|}{Years of study} & \multicolumn{3}{c|}{Gender}  & \multicolumn{5}{c}{Race}\\ \hline \hline 

\multicolumn{1}{c|}{$N$=27} & \multicolumn{1}{c|}{Soph}  & \multicolumn{1}{c|}{Jr}  & \multicolumn{1}{c|}{Sr}   & \multicolumn{1}{c|}{Man}      & \multicolumn{1}{c|}{Woman}  & \multicolumn{1}{c|}{Non-binary}    & \multicolumn{1}{c|}{White} & \multicolumn{1}{c|}{Hispanic} & \multicolumn{1}{c|}{African American} & \multicolumn{1}{c|}{Asian} & \multicolumn{1}{c}{Biracial}                                   \\ \hline 
\multicolumn{1}{c|}{From 8 different institutions} & \multicolumn{1}{c|}{4}      & \multicolumn{1}{c|}{8}  & \multicolumn{1}{c|}{15}   & \multicolumn{1}{c|}{19}      & \multicolumn{1}{c|}{6}  & \multicolumn{1}{c|}{2}    & \multicolumn{1}{c|}{17} & \multicolumn{1}{c|}{3} & \multicolumn{1}{c|}{1} & \multicolumn{1}{c|}{3} & \multicolumn{1}{c}{3}                                   \\ 
\end{tabular}
\end{table*}

\begin{table}[tbh]
  \begin{center}
    \caption{Subfield interests among participants \label{tab: subfield}. Biomedical physics includes biophysics, medical physics, and nuclear medicine. Astrophysics includes asttronomy and astrophysics.}
    \label{tab:subfield}
    \begin{ruledtabular}
    \begin{tabular}{l|l}

 \makecell[l]{Subfield of physics}  &  \makecell[c]{Total}\\ \hline\hline

\makecell[l]{Acoustics}  & \makecell[c]{1}\\ 
 \makecell[l]{Astrophysics}  & \makecell[c]{11}\\ 
 \makecell[l]{Biomedical physics}   & \makecell[c]{6}\\ 
 \makecell[l]{Condensed matter physics} & \makecell[c]{1} \\ 
\makecell[l]{High energy physics}  & \makecell[c]{2}\\
 \makecell[l]{Nuclear physics} & \makecell[c]{1}\\
 \makecell[l]{Optics}  & \makecell[c]{1}\\
 \makecell[l]{Physics education} & \makecell[c]{3}\\
 \makecell[l]{Quantum information}  & \makecell[c]{1} \\ 
  \makecell[l]{Total}  & \makecell[c]{27} \\
 \end{tabular}
 \end{ruledtabular}
  \end{center}
\end{table}












We obtained consent from the interviewees and conducted video-recorded interviews using Zoom. The interviews typically lasted between 60 and 90 minutes. Most participants received a \$10 gift card as a participation incentive. However, interviews that were part of a longitudinal study received a \$20 incentive to recognize the effort of continued involvement the study \cite{alaee2022impact}. 

\subsection{Coding and Analysis}
We aimed to understand the different ways students perceive and understand different subfields of physics. We analyzed the semi-structured interview transcripts using the Dedoose software, which is a tool designed for qualitative data analysis to identify distinct categories~\cite{noauthor_dedoose_2018}. 
The initial step in the data analysis process was deductive coding, which involved identifying and labeling SCCT-related excerpts within data. This step involved assigning specific codes representing different SCCT constructs, such as \textit{learning experiences}. Within each SCCT construct we then used a thematic approach to identify patterns and subcategories. For instance, one emergent subcategory for \textit{learning experiences} focused on the variety of \textit{experimental experiences} students had during their undergraduate studies.


Throughout the coding process, we utilized several methods to provide evidence for the validity and reliability of the coding scheme. 
Initially, one researcher (P.C.) developed the first-pass coding scheme and generated the codebook to identify SCCT categories and subcategories during the first data collection phase. Subsequently, another researcher (R.B.) refined the codebook and applied the same scheme during the second round of data collection. A third researcher (D.Z.) then used the coding scheme on random sections in our data in order to establish inter rater reliability testing (IRR). These codes were further discussed, and revised in weekly meetings with (B.Z.). We continued to discuss and refine the codebook until all disagreements were resolved.


\section{Results}
The results are divided into two subsections. In the first part, we present the subcategories discussed by students and categorize them according to the relevant categories of SCCT. These results align with the first stage of data analysis, which involved a priori coding based on the SCCT categories followed by a thematic grouping within each category. This approach allowed us to gain a comprehensive understanding of the students' experiences within different physics subfields, as well as how these aligned with the SCCT constructs (Sec.~\ref{Results 1}). 
In the second part (Sec.~\ref{Results 2}), we use the results of the SCCT deductive coding and thematic inductive analysis to compare students' perceptions and experiences within two physics subfields, specifically biomedical physics and astrophysics.


\subsection{Results on Research Question 1: Understanding Interest Formation within SCCT Constructs}
\label{Results 1}
This section focuses on research question 1, presenting the themes and findings related to four SCCT constructs: learning experiences, self-efficacy, outcome expectations, and proximal environmental influences. Within each SCCT construct, we organized the findings according to the emergent subcategories that captured the diversity of student responses. It is important to note that not all students mentioned every subcategory. We wanted the subcategories to span the range of ideas expressed in students' decision-making regarding subfield choices.
\subsubsection{Learning experiences and subfield choice}
\begin{figure}[b]
\centering
\includegraphics [trim=3 2 3 7,clip,width=89mm]{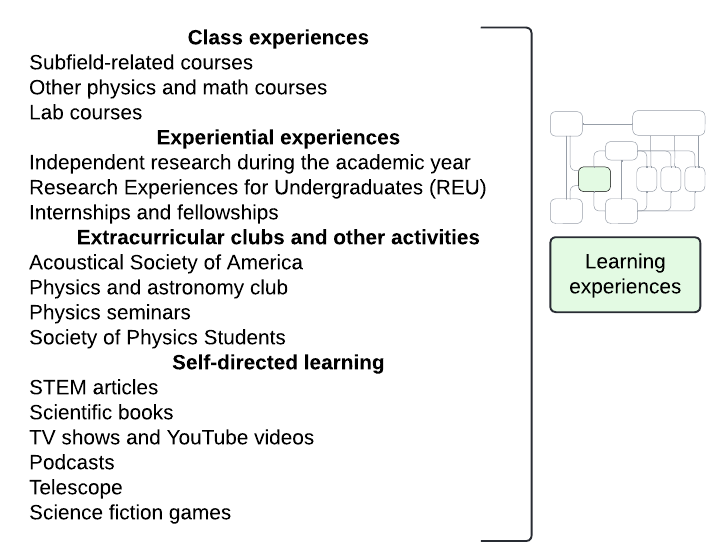}
\centering
\caption{The learning experiences expressed by undergraduate students related to their desired subfield of physics.}
\label{Learning experiences}
\end{figure}
As expected, \textit{learning experiences} played an important role in fostering students' interest in specific subfields of physics. Fig.~\ref{Learning_experiences} shows factors from the interview data that represent the diverse learning opportunities during undergraduate studies. For example, under the SCCT construct of \textit{Learning experiences}, we identified subcategories such as class experiences, research experiences, extracurricular clubs, and self-directed learning.
In the following sections, we will delve into each subcategory of learning experiences, emphasizing their importance in students' interest in their chosen subfields.
\subparagraph{Class experiences.} Taking classes was the most common way to shape students understanding and perspective of specific subfields within physics. These class experiences emerged from data when students reflected upon courses they have completed at the university level, which enriched their positive view of a favored subfield. Class experiences may include one or a combination of courses focused on specific subfields, along with core courses, including lab courses, essential for all physics majors working towards a Bachelor's degree.

Subfield-specific class experiences were the most frequent subcategories that could enhance students' confidence and foster their interest in specific areas of physics. For example, Brett, who was interested in physics education, took many pedagogy classes as part of his Bachelor's program, which included a teaching certification. These learning experiences offered by his institution influenced his interest and confidence to become a high school physics teacher after he graduates. He said, ``I know a lot about teaching... I have already had experience. I was scared when I first started that very first education class I took. I did not know if I [would] be a very good teacher. But the more and more I have [done], the more I have gained confidence in teaching. I do feel really confident about it''~(Brett, Physics education). 

Physics and math courses not specifically tied to the subfield can serve as foundational knowledge and provide relevant skills and concepts necessary for future work. For instance, Mike was interested in quantum information, and coursework in related subjects, such as classical mechanics and computational physics gave him knowledge and confidence. He said, ``I have both the [experiences]; computer science background, physics background, and computational physics. I think pretty confident''~(Mike, Quantum information). Similarly, Thomas, who is interested in particle physics and currently engages in research involving extensive coding in granular materials, expressed, ``I do a lot of math classes and those help with the coding side and kind of the statistical mechanics... There's more than I understood the papers that we looked at once I've taken statistical mechanics. Because we talked a lot about probability distributions when it comes to where particles go and things like that coding in my math classes. I'm in two different stats classes right now and both of them require a ton of Python. In the soft matter research that I do I use Interactive Data Language (IDL),... A lot of code. But just getting the experience of problem solving using code. I can definitely apply those concepts and I understand better''~(Thomas, Particle physics). Thomas saw these skills as laying the foundation for future work in particle physics. 


Labs and experimental physics courses, such as advanced labs, have played a crucial role in shaping students' beliefs that they have a solid foundation in physics principles, problem-solving, and computational skills, enabling them to pursue their subfields of interest. For instance, Frieda stated, ``The habits that I learned last year in advanced lab, those kinds of skills have been really useful as a foundation for being a good scientist in general, in terms of keeping good records and having reliable results.'' She also mentioned learning to use ``all kinds of different [tools] that were specific to those experiments but also often used in other places and foundational to our modern technology. So I think it was worth it''~(Frieda, Astrophysics).

\subparagraph{Experiential learning.} 
Experiential learning opportunities offer hands-on scientific practices and deep engagement in focused areas through faculty-led projects, summer programs, and internships. Bruce emphasized the importance of depth in research projects, highlighting how that allowed him to independently tackle challenges and develop a comprehensive understanding of specific topics ``without immediate instructor assistance.'' Through his research project during the academic year , he noted that he ``fully understands something specifically and very well. He said, ``To understand, to the point where I would understand it inside and out, I think that happened with my research project''~(Bruce, Condensed matter). His statement about understanding his research project ``inside and out'' reflects his self-assurance and motivation, key components of high self-efficacy, which then deepened Bruce's interest in his subfield.
Moveover, experiential experiences provide many benefits for students, fostering personal and professional growth, as Helen experienced it throughout the last year of her junior year research opportunity. She said, ``Because I had the ability to travel and present my physics research and meet other students doing research and I think that [helped] my confidence grow as a physicist''~(Helen, Nuclear medicine). As Helen's quote suggests, research experiences can also provide students with a sense of community and belonging. Interacting with peers who share similar interests can be empowering, as demonstrated by Sophia's positive experience in her biophysics research lab. Sophia expressed her positive experience and sense of connection in her research lab, where she had the opportunity to collaborate with other undergraduate and graduate students. She said, ``My research mentor had about 50-50 women [in her research lab]. One of the students was partially mobile, he's in a wheelchair. Nobody thought that was weird. I was working with other people who look like me. That was pretty sweet. This is probably why I feel welcome and comfortable. I felt welcome as well as safe. Most of the time, I think I feel really confident in doing the experimental work. I know I don't know everything, but I know how to ask. I know how to find people who does.'' By interacting with peers who have similar values, interests, and experiences, Sophia found a supportive community that understood and appreciated her work. Feeling welcome and comfortable is an important factor driving interest and potential long-term involvement for her. She said, ``I hope that getting a bunch of experience and then looking for graduate school. I'm be expecting to get a bunch of experience in it and I can go find job that are like, you have done all of these things, you will be a good fit''~(Sophia, Biophysics).

Further, research mentors have a significant impact on students' subfield choices. Their guidance, shared experiences, passion, and insights into various subfield areas inspire students, fostering passion and motivation to learn more about a particular subfield. For instance, Thomas said, ``I've been working in granular materials for a while, but every time I've learned about anything in particle physics, or accelerator physics, I just really wanted to learn about it... The thing that drew me to the granular materials work was my research mentor. He has this incredible energy about him. I vibe with him well. I think our personalities mesh pretty well. We are both very energetic and optimistic, so I just really wanted to work with him. Then from working with him and experiencing the energy he gives to a room for so long. I have been drawn more and more into the work''~(Thomas, Particle physics).

Experimental learning experiences play a important role in shaping students' interests within various subfields of physics. Most students who discussed research experiences were entering their final year of their bachelor's program, with fewer juniors, sophomores, and one freshman contributing. This highlights a growing interest through research as students progress through their undergraduate studies, emphasizing the importance of early exposure to stimulate curiosity and shape interests in specific physics subfields. These findings underscore the of opportunities that engage undergraduates in research early in their academic careers. By doing so, universities and colleges can nurture students' interests, encourage exploration of diverse subfields, and lay the foundation for future contributions to the field of physics. 

Beyond that, internships also have a positive impact on students' subfield interests. For instance, Sean stated that his internship experience not only provided practical skills but also helped him gain a foothold in the field, boosting his confidence and skills considerably. He said, ``When I came into college, not as confident nowhere near as confident as I am now. I'm still definitely building confidence with it. I my coding experience before college was very low [but] not non existent. I did have an internship working as a data analyst. So that kind of got my foot in the door.'' He continued that ``after working on computational physics he really got a taste for using code consistently across different classes''~(Sean, Astrophysics). Sean's internship experience boosted his confidence in coding skills, paving the way for his deeper engagement in computational physics during college.


\subparagraph{Extracurricular clubs and other activities.} 
\label{Extracurricular clubs}
Extracurricular clubs and activities are where students discuss how their engagement in non-classroom, non-research activities influences their interest in specific subfields of physics. The key focus is the sense of community formed among peers and their active participation in enjoyable activities, including outreach initiatives related to the subfield. Kirk, who was interested in biophysics, found inspiration and ideas by interacting with different students through his participation in the SPS (Society of Physics Students) chapter at his home institution. Caleb, a physics and chemistry club member interested in biophysics, engaged in fun experimental projects during club meetings such as making a ping pong cannon that shot the ping pong ball and synthesizing ibuprofen just starting from a benzene ring. This hands-on experience not only linked his interests in chemistry but also connected him to the biomedical subfield of physics. 
 
Similarly, Grace, interested in astrophysics, shared her involvement in the physics club at her institution, where they organize various events to build a sense of community. These events include watching physics videos together, participating in physics trivia, and engaging in other fun activities to build community among members. In addition, the club also hosts open houses and astronomy nights, providing Grace with the opportunity to work on the rooftop and set up the telescope for the public to look at the sky and engage in scientific discussions. Physics related extracurricular clubs and activities were less commonly discussed compared to research experiences and coursework among the students interviewed. This could indicate a tendency for students to prioritize academic experiences over extracurricular involvement or a lack of opportunities. However, for those students who did participate in physics-related extracurriculars, there were positive impacts on interests within the field, fostering community, and furthering practical applications of physics concepts. Those who engaged in informal science education as part of their extracurricular activities saw gains in knowledge and recognition, which boosts physics identity formation~\cite{hazari_connecting_2010}. 

\subparagraph{Self-directed learning.} Students interested in particular physics may engage in a self-driven approach where they independently seek resources like books, online materials, academic journals, YouTube videos, and social media science communicators to explore their interests in a specific subfield. It allows students to delve deeper into their chosen subfield beyond formal academic structures such as classrooms, improving their understanding and staying updated on the latest news and developments. Sean, for example, expressed his fascination with astrophysics and his proactive approach to learning. To deepen his understanding of astrophysics, he diligently search for any available material, such as instructional guides, books, or other reference materials that designed to provide detailed explanations.
Self-directed learning often begins before college, unlike other learning experiences that are typically inaccessible before higher education. For instance, Beth shared her interest in astrophysics and her reading habits. She said when she was a kid, ``We had books from black holes and stuff that I would read, and then my mom would yell at me to go to bed it is two in the morning''~(Beth, Astrophysics).

\subsubsection{Self-efficacy and subfield choice.}
Self-efficacy refers to an individual's perception of their ability to perform and accomplish a task. To gain a deeper insight into students' self-efficacy, we looked at it through two perspectives: First, we explored the various sources that contribute to the development of self-efficacy among students, and second, we investigated the benefits associated with having a higher level of self-efficacy in relation to making choices within a specific subfield. Fig.~\ref{Self_efficacy} shows the self-efficacy factors that emerged from interview data.
\begin{figure}[h]
\centering
\includegraphics [trim=0 0 0 0,clip,width=89mm]{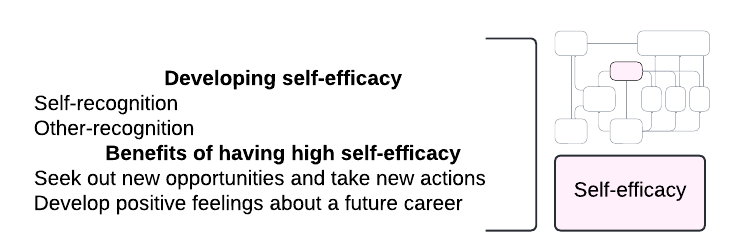}
\centering
\caption{Self-efficacy expressed by physics majors.}
\label{Self-efficacy}
\end{figure}
\\
\paragraph{\textbf{Developing self-efficacy.}}
Students with limited learning experiences often felt less confident in their abilities to pursue their desired subfield. However, as they learned more about their chosen area, their self-efficacy, or belief in their ability to succeed, changed. This transformation was influenced by both self-recognition, where  students saw their own growth and potential, and other-recognition, where they were acknowledged and praised by peers or mentors. This cycle of recognition boosted their motivation to learn, engage in research, and pursue their desired subfield in physics. 

\subparagraph{Self-recognition.}
Our data shows a link between students' self-efficacy and their past achievements and learning experiences. According to SCCT, self-efficacy should also influence interests in specific physics subfields. A student's successful academic performances, such as achieving high grades, can lead to a heightened self-recognition and investment in their learning, motivating them to pursue related opportunities and specific physics subfields.
One student, Emma, exemplified this relationship between past successful performance and self-efficacy. She said, I am pretty sure I will get accepted [in grad school in physics education], I have good grades and a good GPA, and I am very clear about what I want... I feel confident about it. I am nervous, but I have worked really hard to prepare [for the] licensing program.'' She added that she might not be fully ready to start teaching and may encounter some challenges, but she feels ``prepared enough to work through those issues''~(Emma, Physics education).

Another example involved Beth, who discussed the significance of the community within her desired subfield and its impact on mental health and work-life balance. She remarked that through conversations with other people and personal observations, it seems that significant progress has been made in [the community] in recent years alone, ``so that makes me feel more confident and being able to enjoy that life''~(Beth, Astrophysics).
Emma's preparedness and anticipation of challenges, as well as Beth's recognition of progress within her academic community, exemplify the importance of self-recognition in navigating academic and professional pursuits. Recognizing one's capabilities and the support of a nurturing environment can empower students to pursue their goals with confidence and resilience.


\subparagraph{Other-recognition.} 
When students were acknowledged and received positive feedback from the scientific community, other students, friends, family, and general public, they became more confident in their abilities to achieve academic and career goals. This recognition can include supportive verbal encouragement (e.g., telling the student that `you can do it', `you did a good job', and `you are a problem solver') and providing opportunities for students to present their work within the scientific community (e.g., sending the student to the conference). For instance, as we mentioned in Sec.~\ref{Extracurricular clubs}, the astronomy club's open houses and astronomy nights allow Grace to set up the telescope on the rooftop, gaining public recognition while facilitating sky viewing and scientific discussions. Moreover, Joshua said, ``My mentor says, `You can do it just like before.' So, my mentor gave me much confidence and erased my worries''~(Joshua, High energy physics). 

Emma, who has an interest in teaching, received additional recognition from her academic advisor, who also leads many of the math education preparation program courses. 
Another of Emma's mentors, who is also a instructor and supervising Emma as a learning assistant, ``constantly pushing [her] to go to graduate school.'' Emma expressed gratitude, stating, ``When I first started at my institution, she was the chair of the physics department and helped me determine which classes I needed to take. She's been a big like support for me, continuing and finishing the [program]. She's like, `you could just get a masters and then teach' so she really wants me to keep doing my own education.  She is definitely been huge encouragement''~(Emma, Physics education). Their guidance has motivated Emma to apply for graduate school and pursue her teaching interest. 
For Beth, the other recognition stemmed from recognizing her work by her mentor. She said, ``Most of my confidence comes from my mentors and any research experience I have done... Because I do research, and then the mentors come up with a lot of good things to say about me and invite me to conferences''~(Beth, Astrophysics). Being invited to present her research findings at conferences not only validates her abilities in conducting research but also provides her with a platform to share her work and be recognized by the scientific community.

\paragraph{\textbf{Benefits of having high self-efficacy.}} Thus far, we explored various factors contributing to developing students' self-efficacy. In the subsequent section, we identify themes related to students' high levels of self-efficacy, which significantly impacted their decision-making process regarding subfield choices and future career paths. This finding aligns with the SCCT diagram, where self-efficacy mediates between learning experiences and developing interest and subsequent choice actions in the chosen subfield. 

\subparagraph{Seek out new opportunities and take new actions.} Having high self-efficacy often leads to seeking new opportunities and taking action~\cite{wigfield2000expectancy}. Students who strongly believe in their capabilities are more motivated to actively seek opportunities to learn more about them. Many students expressed their willingness to explore more research positions and engage in new learning experiences. For example, Mike ``is researching with a professor to kind of dip his toes into the field of quantum information''~(Mike, Quantum information). 

Bruce is confident in his ability to conduct experimental research in condensed matter physics in the future because he possesses the fundamental skills required;``I have the basic tools to be able to understand how do something [work], I know how to look up new tools. I know how to figure out how the tool works. I know how to try to set it up and try to troubleshooting, I think that's kind of the foundation for experimental work and I have kind of a precursor idea of how to have enough of a physics background to be able to understand what goal actually want out of the experiment. So, I do feel confident.'' Bruce's confidence in his experimental skills and understanding of physics provides a strong foundation for his future research in the field of condensed matter physics. He commented, ``The idea of doing condensed matter research, or some research that is kind of an extension of what I'm doing right now and that's all experimental''~(Bruce, Condensed matter physics). Linking this to his earlier emphasis on fundamental skills, Bruce's high self-efficacy motivates him to seek new opportunities and engage in condensed matter physics research.


\subparagraph{Develop positive feelings about a future career in a chosen subfield.} 
Students with high self-efficacy expressed confidence and positive feelings regarding their future career paths. Emotional factors, physics identity, and a sense of belonging are not explicitly captured in the SCCT map. However, the relationship between self-efficacy and a sense of belonging is mutually influential~\cite{alaee2022impact}. Pursuing an area of interest brings satisfaction, happiness, and a sense of fulfillment. For example, Thomas expressed his confidence and positive feelings about the community in his chosen subfield of high energy physics. He said, ``I am confident. I like the community. I have worked with a few different professors, and we have a research group. I have talked to a bunch of people from different places and have really liked it so far. The work is fun''~(Thomas, Particle physics). 

Some of our students have high confidence and positive emotions which enhance their interest in their chosen subfield of physics. For instance, Max, shared that 
``I probably could do well in it. I think I have the skills necessary to go into the subfield. I feel physics education is what suits me really well''~(Max, Physics education). Max is confident that he has the skills to excel in physics education, strengthens his interest in physics education. This reinforces his belief in his ability to succeed.

\subsubsection{Outcome expectations and subfield choice}
When students contemplate their future careers within their chosen subfield, considering their outcome expectations becomes very important. Outcome expectations capture the envisioned results or achievements students anticipate in their future careers within a subfield. These expectations can vary widely among students, including factors such as job security, maintaining a healthy work-life balance, and achieving personal fulfillment. 
Fig.~\ref{Outcome_expectations} offers a glimpse into the common outcome expectations expressed by students across various physics subfields. In the following sections, we categorize these outcome expectations into three main groups:
exploring career paths, skill development and expertise acquisition, and integrating work and personal life.

\begin{figure}[t]
\centering
\includegraphics [trim=0 0 0 0,clip,width=89mm]{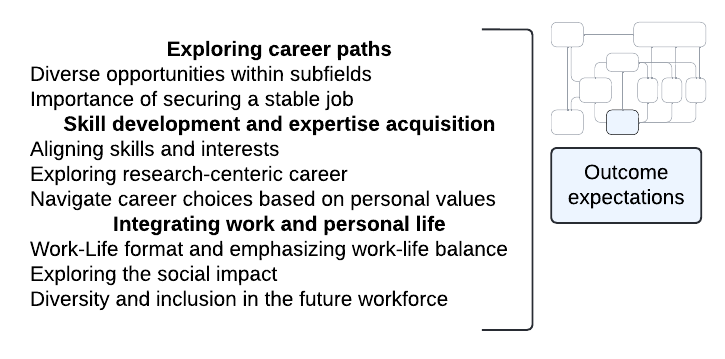}
\centering
\caption{Outcome expectations expressed by undergraduate students.}
\label{Outcome expectations}
\end{figure}

\paragraph{\textbf{Exploring career paths: navigating diverse opportunities and prioritizing stability}} refers to the process of investigating and considering various options for students' future professional paths.
\subparagraph{Diverse opportunities within subfields.} 
Across different subfields, students expressed their awareness of the wide range of opportunities available to them due to the nature of the subfield. Sophia, for example, discussed her ideas of pursuing a career in the field of biophysics. She said,``I could be working in a variety of things. I could see being doing medical testing. I could [be] in a pharmacy. I could see doing biotech, [it] is really big... I can do data processing stuff. So ideally, I would be working in a lab, trying to do medical research and poking at stuff, but also I think there are a lot of good physics options for these based on the interest in biophysics that I have and that I have worked in''~(Sophia, Biophysics). 

We also noted that students with high self-efficacy were more open to exploring different career opportunities within their chosen subfield of physics. David, for example, has explored ``companies that are doing sonar research as well as different national labs that are doing research in acoustics and universities where I could potentially [pursue] a faculty position. I do not know where I would end up going, but I could see myself going any of those three routes''~(David, Acoustics).


\subparagraph{Importance of securing a stable job.} One outcome expectation expressed by several students regarding their preferred subfields is the assurance of obtaining a good job. While it is important to note that financial concerns were not the only factor in choosing a subfield, it is worth considering the outcome expectation of a good income in particular subfields. For example, Helen acknowledged that while money is not the primary motivator, money can enable her to contribute to society in other ways~(Helen, Nuclear medicine). On the other hand, David expressed satisfaction in finding a field he enjoys and recognized its favorable salaries based on published statistics. 
The expectation of good financial and job security was not restricted to applied subfields. Being a ``practical person,'' Frieda emphasized positive beliefs about financial and job security in her pursuit of Astrophysics. These students view income and jobs as a motivating factor in their subfield choices, giving them a sense of security and stability.

\paragraph{\textbf{Developing skills and expertise}} refers to the expectations of acquiring specific skills and tasks crucial for daily work within a chosen subfield. Immersing oneself in learning experiences within the subfield offers valuable expertise and specialized skills. Students anticipate gaining a profound understanding of the subject matter, conducting thorough research, and ultimately becoming experts in their chosen field.
\subparagraph{Aligning skills and interests.}
Considering the subfield characteristics helps students align their skills, interests, and values with the subfield that suits them best. This is important for students as it ensures that they not only excel academically but also find fulfillment and satisfaction in their chosen subfield of physics. By understanding the unique nature and requirements of different subfields, students can make informed decisions about their career paths. For instance, Dalton emphasized that the best part of the Astrophysics field involves employing ``beefy mathematics to derive equations and analyzing observational evidence''~(Dalton, Astrophysics).

Kirk's explanation of why he enjoys biophysics reflects his enthusiasm for the nature and scale of the subfield. He expresses, ``The reason I like biophysics is because... I love the environment.'' In his mind, biophysics involves more than just ``observing through a microscope and taking measurements.'' He said, ``In reality, I think it is like taking videos of things moving at cell scale, and I know some people measure that kind of motion. So like, taking the videos and analyzing it seems really cool to me''(Kirk, Biophysics). He finds the idea of visualizing these motions fascinating, even though they may not be visible to the naked eye, ``Because you're actually seeing something move. Like obviously, I'm not seeing it with my eyes, but you're still seeing it.'' This ability to visualize and analyze dynamic processes at a microscopic level clearly captivates Kirk's curiosity and imagination. Kirk also shares his dislike for chemistry, citing the heavy burden of memorization as a drawback. He contrasts this with physics, where he finds joy in the problem-solving aspect. For Kirk, physics offers a blend of understanding through problem-solving rather than rote memorization, making it a more engaging and enjoyable subject for him.

\subparagraph{Exploring research-centric career.} Some students anticipate immersing themselves in research in their future careers, driven by the desire to discovery and make meaningful contributions to their field. For instance, Adam provided examples of research-oriented career paths within nuclear physics. He mentioned possibly being a ``university researcher who is just doing research''~(Adam, Nuclear physics). In addition, Adam emphasized that the outcome expectation extends beyond exploring new knowledge, as it also includes sharing findings through publications with others.

\subparagraph{Navigating career choices based on personal values.} Another common outcome expectation for students pursuing subfields of physics with a direct impact on people and society (e.g., medical physics, biophysics, and physics education), is their eagerness to contribute to the well-being of others and create a positive societal influence.  
For instance, Cale expressed his perspective on astrophysics, acknowledging ``it's interesting, but at the same time, is it realistic?'' and questioning its practicality and tangible impact. Conversely, he highlights the benefits of medical physics, emphasizing the potential to ``help people'' in need. Nuclear physics also holds promise for advancements in medical scanning technologies, albeit he emphasized that ``Not too much a fan of the weaponization part of it, but it is a double-edge sword''~(Cale, Medical physics).  This highlights how students think carefully as they explore physics careers, balancing their curiosity with how they can make a difference, and staying true to their values.\\

\paragraph{\textbf{Integrating work and personal life}} refers to outcome expectations that focus on the overall lifestyle connected with a subfield, rather than considerations of the day-to-day work routine.
\subparagraph{Work-life format and emphasizing work-life balance.}
Some students recognized that some careers within physics might offer opportunities to travel to conferences and present their work. For instance, Beth mentioned that she ``could hopefully be able to have some kind of work-life balance... If I can travel to conferences to do work, I would love that''~(Beth, Astrophysics). We also found that some students are influenced by their mentors and the way they approach their work. Bruce, for example, mentioned that his mentor demonstrated a balanced way of doing research, which had a significant impact on him. He said, ``a sense of how he does research... He's not always super stressed about everything. He's not overworking himself. This idea that you can be a researcher and not overwork yourself all the time is also quite enticing''~(Bruce, Condensed matter physics). It led him to realize that being a researcher does not necessarily mean constantly overworking oneself, but that there is a way to maintain a balance between work and personal life. This realization has given Bruce a new perspective on pursuing a career in condensed matter physics. 
 \subparagraph{Finding community at work.} In two different ways, students view social interactions as a significant outcome expectation in their future careers within their chosen subfields.
 
\textit{{\textbf{Desire to avoid isolation.}}} Some students emphasized that their personal goals for pursuing a specific type of physics typically align with the nature of their desired subfield. Max, for instance, expressed his preference for being in a subfield that offers social connection and engagement with people. He said he is not looking at careers where he is ``just sitting in a lab by myself, social is important to me.'' He is a part of social groups at his institution and ``I've always had a lot of very close [friends from other group], which makes me want to do like something more social and more extroverted''~(Max, Physics education).  

\textit{{\textbf{Diverse teamwork.}}} Additionally, students emphasized the value of diverse teamwork in their future careers. Sean expressed the importance of ``spending time with people who are [highly] focused on astrophysics,'' allowing for deeper engagement and learning~(Sean, Astrophysics). Similarly, Helen found the idea of medical physics intriguing because of the physics aspect and because it provides opportunities to interact with professionals from ``different fields such as physicians, patients, biologists, computer scientists, and engineers.'' She stated that this collaboration aspect allows for a broader perspective and the exchange of knowledge and ideas within the STEM-related fields, ``You are working with not just other physicists, you are working with a bunch of other people in STEM-related fields''~(Helen, Nuclear medicine). 

\subparagraph{Diversity and inclusion in the future workforce.} Some students' perspectives shed light on how anticipated cultural factors can significantly influence their choices within their subfield of interest. For instance, Grace said, ``I've been thinking a lot more on working as like faculty. I'm not sure about [working in] a company. I'd have to do a little more research into that and what the atmosphere there is like. I feel pretty strong [about pursuing astrophysics]. I think I feel physics and astronomy, particularly, have been very male-dominated fields. In many astronomy classes, there is a notable focus on the historical context, which often highlights the contributions of male figures who have received Nobel Prizes... But this is the woman that did all the work. So, I feel pursuing that specifically, at a historically women's college is pretty strengthening. I like it. I see myself as a physicist''~(Grace, Astrophysics). Although Grace noticed these negative elements in the male-dominated culture and historical focus on male achievements in physics and astronomy, she still wanted to pursue astrophysics. She saw herself as a physicist and felt that pursuing this field at a historically women's college was empowering and aligns with her passion. She also appreciated the supportive environment where her contributions were valued and was inspired to advocate for gender diversity in the field.

\subsubsection{Proximal environmental influences and subfield interest}
\label{Proximal environmental influences and subfield interest}
During interviews, students discussed how environmental factors in their institution and personal lives could impact their subfield choices. The analysis of these factors revealed a range of influences and barriers that students considered when making decisions about their subfield of interest. The environmental influences discussed by students are shown in Fig.~\ref{Proximal_environmental_influences}. 

\begin{figure}[h]
\centering
\includegraphics [trim=11 3 6 7,clip,width=80mm]{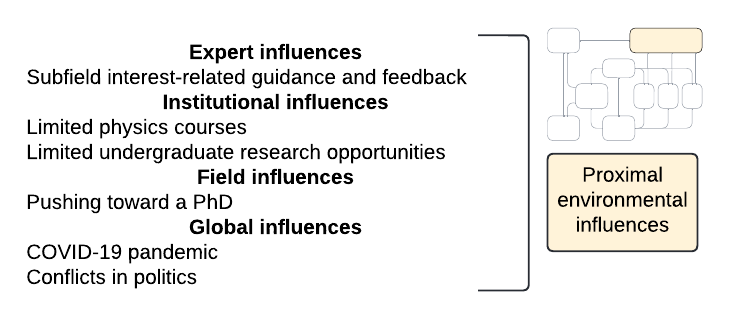}
\centering
\caption{Proximal environmental factors expressed by undergraduate students.}
\label{Proximal environmental influences}
\end{figure}
\subparagraph{Expert influences.} 
The influence of expert people, particularly professors, and mentors, can play a significant role in shaping students' decisions regarding their subfield choices. Joshua emphasized his mentor's qualities of kindness and effective leadership, which had an impact on his academic journey to start a high energy physics research with him.

\subparagraph{Institutional influences.} Students also discussed the impact of institutional factors within their educational environments. Barriers such as a lack of support from the institution, limited coursework, or limited research opportunities in particular subfields were identified as potential challenges that could affect their subfield choices. For example, Frieda said, ``We do not have opportunities for astrophysics research at my school, and so I have not had the opportunity to get involved yet, and if we had classes in astrophysics at [my school], I would 100\% take them, but we do not''~(Frieda, Astrophysics). Leo is concerned about the ``shortage of medical physicists'' at his institution. He said, ``Nobody knows about the field because there is no outreach. It is not like astrophysics or condensed matter physics that most people learn about it during the last couple of years of undergrad''~(Leo, Medical physics). The academic culture and institutional resources available can profoundly influence students' interest in chosen subfield and their career paths. 
\subparagraph{Field influences.}
Choosing to major in physics shapes a student's future career choices. Some may feel burdened by the culture of the field values some career paths more than others and students pick up on which fields receive respect. This influence is clear in Brett's experience, who thought about getting a master's in education, perhaps while teaching high school or at some other point in his career. However, he noted, ``I've always kind of wanted a PhD, just to have the doctor [title]... In physics, it's very much like being pushed into a PhD track~''(Brett, Physics Education).

\subparagraph{Global influences.} External factors like the COVID-19 pandemic have influenced students' choices concerning their desired subfields and future career trajectories. Cale, for instance, shared his perspective on how the COVID-19 situation has ``pushed him towards medical physics a bit''~(Cale, Medical physics). Cale emphasized the challenges the medical field faces, especially in the United States, and the pressing need for more support and human resources to address its broken aspects. The pandemic has heightened the stresses within the healthcare system, reinforcing Cale's belief that he can contribute and make a meaningful impact in this field.


\section{Results on Research Question 2: Comparing Interest Formation}
\label{Results 2}
The analysis of Research Question 2 focused on exploring the differences in how interest is formed between two specific subfields: biomedical physics (including biophysics, medical physics, and nuclear medicine) and astrophysics. Studying these differences provided valuable insights into the factors that shape interest in various subfields. We observed many similarities in self-efficacy and proximal environmental influences between those students interested in biomedical physics and in astrophysics. However, one proximal environmental factor stood out distinctly. As mentioned in Sec.~\ref{Proximal environmental influences and subfield interest}, the COVID-19 pandemic was linked to a higher interest in biomedical physics. Both astronomy and biomedical physics faced challenges with limited access to coursework.

Additionally, there were significant differences between how often students mentioned particular learning experiences or outcomes expectations. Table~\ref{tab:Comaprison} illustrates the differences in learning experiences and outcome expectations between astrophysics and biomedical physics subfields, which we review in the following subsetions.
\subsection{Exploring Learning Experiences: Biomedical Physics versus Astrophysics}
When comparing the learning experiences in astrophysics and biomedical physics distinct patterns can be observed.
The popularity of astrophysics can be attributed to four main factors: its portrayal in popular culture and popular science media, exposure in high school and college courses, and strong inherent interest, and the fact that people can look up and see the moon, stars, planets, sun, and use telescopes. Popular science inspires students' interest in astrophysics. Beth pointed out, ``Because astronomy is really big in our culture, I think everyone has heard of astronomy at one point''~(Beth, Astrophysics). Some students mentioned that their exposure to astrophysics began in high school and even before that, with supportive high school teachers fostering their interest in astronomy. Finally, many universities offered college classes in astronomy and astrophysics, which allowed students to gain a deeper understanding of the field, solidifying their interest and expanding their knowledge. Dalton said, ``I absolutely adore [astronomy]. It is loads of fun.'' For him, the experience is enjoyable and less focused on equations, allowing him to ``just get to enjoy learning about how things roughly happen''~(Dalton, Astrophysics). 

In contrast to astrophysics, biomedical physics was not commonly encountered through popular science media or high school or college coursework. Instead, interest in biomedical physics was primarily influenced by engaging in research and practical experiences within the medical field. So, while astrophysics often captured students' attention before college through its presence in popular culture and in college through coursework, the formation of interest in biophysics relied more heavily on research opportunities and other observational and participatory activities, which typically did not happen until the later stages of an undergraduate degree. Caleb, for example, found great value in job shadowing doctors, which provided him with firsthand insights into the hospital environment and further fueled his interest in biomedical physics.

\subsection{Exploring Outcome Expectations: Biomedical Physics versus Astrophysics}
Students interested in astrophysics are driven by the opportunity to engage in research on a bigger scale and contribute to discoveries within the field. Sean expressed a desire to experience discovering something and leaving an impact~(Sean, Astrophysics). Moreover, multiple students were confident of finding a good job in astrophysics-related careers. 

In contrast, students pursuing careers in biomedical physics prioritized making a positive impact on others and society as a desired outcome expectation. They were motivated by the prospect of using their scientific knowledge and skills to improve the quality of life for others. Helen emphasized the significance of ``having short-term effects on whom you impact. [Look] back on a list of patients and think about all the people you have impacted and helped their families and like save their lives''~(Helen, Nuclear medicine). 

\begin{table*}[tbh]
\caption{Comparing SCCT Constructs: Astrophysics and Biophysics Subfields.}
\begin{tabular}{c|l|l}
\textbf{\makecell[l]{Subfield of Physics/\\SCCT Constructs}}  &  \textbf{\makecell[c]{Astrophysics} } &  \textbf{\makecell[c]{Biomedical Physics}} \\
 \hline

{\textit{Learning Experiences}}	&\makecell[l]{Gain exposure through popular science media} &\makecell[l]{Engage in research on a larger scale}\\
{\textit{}}	&\makecell[l]{High school and college classes} &\makecell[l]{} \\
{\textit{}}	&\makecell[l]{Strong inherent interest} \\
{\textit{}}	&\makecell[l]{Visibility and Accessibility} \\
\hline

{\textit{Outcome Expectations}}	&\makecell[l]{Secure fulfilling careers} &\makecell[l]{Make an impact on others and society}\\
{\textit{}}	&\makecell[l]{Contribute to discoveries} &\makecell[l]{Enhance the overall quality of life for society} \\


\end{tabular}
  \label{tab:Comaprison}
\end{table*}

\section{Discussion and Conclusions}
Our research question 1 (RQ1) aimed to explore how Social Cognitive Career Theory (SCCT) constructs such as learning experiences, self-efficacy, outcome expectations, and  proximal environmental influences affected students' interests in different subfields of physics. Our findings revealed that students gained knowledge in diverse ways, drawing from class experiences, research opportunities, extracurricular activities, and self-directed learning. 

Moreover, our exploration of self-efficacy and its role in subfield choice revealed that students' beliefs in their abilities were influenced by both self-recognition and other-recognition, which aligns with other physics education research on self-efficacy~\cite{hazari_connecting_2010, li2020perception, kalender2019female}. For instance, facilitating extracurricular informal science education activities lead to recognition that boosted self-efficacy and identity formation as participants treated the facilitators as experts, even though they had not completed their undergraduate degree. Students, such as Grace, who engaged in informal science education highlighted the link between self-efficacy and recognition by self and others. Our findings align with those of thePrefontaine \textit{et al}.~\cite{prefontaine2021informal} study, which highlighted the unique role of informal teaching and learning programs in shaping students' interests in physics. Such programs offer opportunities for both participants and facilitators to interact in a way that fosters the development of a strong physics identity.

Furthermore, our study highlighted the importance of outcome expectations, which refer to students' perceptions about the potential outcomes of their career decisions, whether those are realistic or not. For example, popular science can provide broad exposure to specific subfields, such as astronomy. However, it often creates unrealistic outcome expectations for students, such as spending lots of time at telescopes rather than doing mostly data analysis and modeling at a computer. Moreover, environmental factors within their institutions and personal lives, such as the availability of courses, may limit or positively influence students' decisions to pursue specific subfields in physics. Physics departments should facilitate opportunities for students to participate in discussions, collaborations, and shared experiences with peers who have similar interests in specific subfields.

While SCCT provided a comprehensive framework for understanding interest formation, our second research question~(RQ2) delved into a comparative analysis of interest formation across two specific subfields: astrophysics and biomedical physics. It is important to acknowledge that these two subfields are the most prominently represented across our data, with 11 participants interested in astrophysics and 6 in biomedical physics. With a broader representation of students across additional subfields, it is possible that additional other intriguing comparisons would emerge, offering further depth and insight into the factors influencing student's interest formation within the field of physics.

Our exploration showed significant differences in how and when students encountered learning experiences relevant to the subfield. Regarding outcome expectations, We found that students who are interested in careers in astrophysics focused on research, discoveries, and securing fulfilling jobs, while those in biomedical physics prioritized impacting society and improving quality of life.

\section{Implications \& Future Research}
To help students explore these subfields, especially early on, physics departments should offer a range of opportunities. These can include elective courses, guest lectures, research experiences, and targeted events focusing on specific subfields. Subfield-relevant examples could also be woven into lecture and laboratory courses that don't align with any specific subfield. Gaining insights into the varying factors that influence interest development across subfields of physics, can help us redesign our undergraduate programs to better support students' awareness of opportunities.

Nevertheless, it is important to note the limitations of our research, such as the relatively small sample size. In some cases, we only had one or two students representing certain subfields, which limits the generalizability of our findings. Future studies could address this limitation by expanding on our work to explore a broader range of subfields and by including a larger and more diverse participant pool, particularly including greater institutional diversity. Additionally, longitudinal studies would provide valuable insights into how students' interests in subfields evolve over the course of their undergraduate program, helping departments to better support the career decisions of their students.

In the SCCT framework, two critical factors are not included: sense of belonging and physics identity. Previous studies have highlighted the significance of these elements in shaping students' interest, retention, and career decisions within the field of physics~\cite{Lock-Physics-2013, Godwin-Identity-2016, hazari_connecting_2010}. 
Our future work aims to integrate these factors alongside the SCCT framework, culminating in the design of an assessment for career decision-making in physics. Recent reports on physics education, such as the Effective Practices for Physics Programs (EP3)~\cite{mckagan_career_2021} and the TEAM-UP report~\cite{national_task}, highlight the importance of fostering career awareness and preparation within physics education. We want to do additional research on other facets of interest formation, such as the methods that students want to pursue (e.g., experimental, computational, theoretical) and how they decide between graduate school or employment after graduating. By examining the influences of various factors on students' decision-making processes, we can enhance our understanding of how students form their identities and make informed choices in physics.

\bibliography{2024_SCCT_paper}
\end{document}